%% file: Lb_JpsiLambda_v3.tex
\newif\ifLHCbLook
\LHCbLooktrue

\ifLHCbLook
  \documentclass[12pt,a4paper]{article}
\else
  \documentclass[aps,prd,reprint,groupedaddress,amsmath,amssymb,nofootinbib,floatfix]{revtex4-2}
\fi
\usepackage[utf8]{inputenc}\usepackage[T1]{fontenc}
\usepackage{amsmath,amssymb}\usepackage{bm}\usepackage{booktabs}\usepackage{array}
\usepackage{amsthm}
\theoremstyle{plain}

\theoremstyle{remark}

\usepackage[colorlinks=true,linkcolor=blue,citecolor=blue,urlcolor=blue]{hyperref}
\usepackage{braket}\usepackage{graphicx}
\usepackage{etoolbox}

\ifLHCbLook\else
\makeatletter
\patchcmd{\doauthor}{\frontmatter@footnote{#2}}{%
  \frontmatter@footnote{These authors contributed equally to this work.}%
  \textsuperscript{,\,}\frontmatter@footnote{#2}%
}{}{\PackageError{authornotes}{Unable to separate author footnotes}{Check the REVTeX author-note definition.}}
\makeatother
\fi

\ifLHCbLook
\input{styles/lhcb-like}
\else
\newcommand{\singlefigwidth}{\columnwidth}
\fi
\newcommand{\Lb}{\ensuremath{{\Lambda_b^0}}}\newcommand{\Jpsi}{\ensuremath{{J/\psi}}}\newcommand{\Lam}{\ensuremath{{\Lambda}}}
\newcommand{\Lbbar}{\ensuremath{{\overline{\Lambda}_b^0}}}

\providecommand{\ap}{}\renewcommand{\ap}{a_+}\newcommand{\am}{a_-}\newcommand{\bp}{b_+}\newcommand{\bmn}{b_-}

\newcommand{\Tr}{\mathrm{Tr}}\newcommand{\dd}{\mathrm{d}}
\newcommand{\helicity}{\ensuremath{\lambda}}

\begin{document}
\title{An exact separability criterion for the \texorpdfstring{$J/\psi\Lambda$}{J/ψΛ} spin state in \texorpdfstring{$\Lambda_b^0\to J/\psi\,\Lambda$}{Λb⁰ → J/ψ Λ} decays at the LHC}
\author{Yujie Li}
\email[Contact author: ]{yujie.li@cern.ch}
\author{Xiaofan Hu}
\email[Contact author: ]{xiaofan.hu@cern.ch}
\author{Liming Zhang}
\email[Contact author: ]{liming.zhang@cern.ch}
\affiliation{Department of Engineering Physics, Tsinghua University, Beijing, China}
\date{17 November 2026}

\begin{abstract}
We derive an exact separability criterion for the spin state of the $J/\psi\Lambda$ pair produced in $\Lambda_b^0\to J/\psi(\to\mu^+\mu^-)\,\Lambda(\to p\pi^-)$ decays, taking into account the $\Lambda_b^0$ polarization at the LHC. 
Within the four-amplitude family $\{a_\pm,b_\pm\}$, the system is separable if and only if $(1+P_b\cos\theta)a_+b_+=(1-P_b\cos\theta)a_-b_-=P_b\sin\theta b_+b_-=0$, where $\theta$ is the $\Lambda$ production angle defined in the LHC environment.
We propose a profiled likelihood ratio test to probe entanglement in the full five-dimensional angular analysis. Idealized pseudoexperiments compare this test with a negativity test and a test based on the angular moments $M_7$ and $M_9$ for rejecting the $a_+=b_-=0$ hypothesis. At both $3\sigma$ and $5\sigma$ significance, the proposed test achieves a $90\%$ rejection power with about half as many signal events as the other two methods. These results support the prospects for entanglement evidence with samples on the existing scale and for observation with larger samples. 
\end{abstract}

\maketitle
\section{Introduction}
\label{sec:intro}
Entanglement~\cite{Schrodinger1935} is one of the sharpest departures of quantum mechanics from classical physics, and probing it at the highest accessible energies has become an active LHC program~\cite{afik2020,afik2021,atlas_topent,cms_topent,atlas_hzz}.
Dedicated frameworks have been developed for the two Higgs-boson decay channels $H\to W^+W^-$~\cite{prd_hww} and $H\to ZZ$~\cite{prd_hzz}, and baryonic systems have been examined at BESIII~\cite{bes3_natphys,bes3_nature}.
Entanglement in baryon--antibaryon final states has also been studied in charmonium decays~\cite{fabbrichesi}.
ATLAS has measured spin entanglement between the two $Z$ bosons in $H\to ZZ^{*}\to4\ell$ decays~\cite{atlas_hzz}, demonstrating the experimental feasibility of this program at the LHC.
In $B$ physics, the published LHCb helicity amplitudes for $B^0\to J/\psi K^*(892)^0$ have been used to infer entanglement between the final-state vector mesons and substantial violation of a Bell inequality within the adopted analysis framework~\cite{Fabbrichesi:2023idl}.

In this paper, we study the decay $\Lb\to\Jpsi(\to\mu^+\mu^-)\,\Lam(\to p\pi^-)$ in the LHC environment. The spin-$\tfrac12$ parent constrains the final-state spin projections, forbidding  the difference of helicity $|\helicity_{\Lam}-\helicity_{\Jpsi}| = \tfrac{3}{2}$ and making this decay a potential probe of quantum entanglement. This constraint is weaker than in the Higgs decays mentioned above, where a scalar initial state forces its decay products into a spin singlet. Nevertheless, particle production in proton--proton collisions, combined with the detector configurations of the LHC experiments, enables measurements of the $\Lb$ polarization~\cite{lhcb_jpsilam,ATLAS:2014swk}.
The two-body final state has a $2\otimes3$ spin Hilbert space, for which the Peres--Horodecki criterion provides a necessary and sufficient test of separability~\cite{peres1996,horodecki1996}. 

References~\cite{lhcb_jpsilam,ATLAS:2014swk} measured the angular distribution of this decay and the four helicity amplitudes permitted by angular-momentum conservation. Both measurements find that the $\Lam$ baryon is dominated by the $\helicity_{\Lam}=-\tfrac{1}{2}$ state and that the $\Lb$ polarization is consistent with zero.
In this paper, we show that entanglement requires a nonzero $\helicity_{\Lam}=+\tfrac{1}{2}$ component---that is, a right-handed $\Lam$---and examine how the choice between fitting schemes with and without the assumption of zero $\Lb$ polarization affects the entanglement measurement.

The paper is organized as follows. Section~\ref{sec:system} constructs the spin-density matrix and Sec.~\ref{sec:ent} establishes the exact separability criterion. Section~\ref{sec:num} rewrites the spin-density matrix in irreducible-tensor form, connects its coefficients to the measured angular moments and identifies those that carry the entanglement information. Section~\ref{sec:angular} introduces the profiled likelihood ratio test proposed in this work, together with the fitted negativity and the $M_7$--$M_9$ moment test, and compares their rejection power on common idealized pseudoexperiments. Section~\ref{sec:data} applies the criterion to the published LHCb angular moments, through the posterior reconstruction of the negativity and the joint $M_7$--$M_9$ test. Section~\ref{sec:concl} presents our conclusions. Appendix~\ref{sec:toy_sensitivity} details the generating points and numerical results of the pseudoexperiments, and Appendix~\ref{app:numbers} collects the remaining numerical results.
We emphasize that this work addresses entanglement rather than Bell nonlocality. A statistically robust test of the latter requires a dedicated treatment of the measurement settings and experimental uncertainties, which we leave to future work.

\section{The \texorpdfstring{$\Lambda_b^0\to J/\psi\,\Lambda$}{Λb⁰ → J/ψ Λ} spin-density matrix}
\label{sec:system}
The $\Lambda_b^0\to J/\psi\,\Lambda$ decay, together with the subsequent decays $\Jpsi \to \mu^+\mu^-$ and $\Lam \to p \pi^-$, is described by five angles: $(\theta, \theta_b, \phi_b, \theta_l, \phi_l)$.
We follow the definitions in Ref.~\cite{lhcb_jpsilam}, restated here for completeness.

The decay angle $\theta$ and the spin projection $m_b$ of the initial-state $\Lb$ are defined with respect to the direction $\hat{n}$, along which the polarization is measured~\cite{Blake:2017une}.
The unit vector is given by \mbox{$\hat{n} = (\vec{p}_{\rm beam} \times \vec{p}_{\Lb})/|\vec{p}_{\rm beam} \times \vec{p}_{\Lb}|$}, where $\vec{p}_{\Lb}$ and $\vec{p}_{\rm beam}$ are vectors in the direction of the $\Lb$ baryon and the beam in the center-of-mass frame of the $pp$ collision; in the case of the LHCb detector, $\vec{p}_{\rm beam}$ points along the beam entering the detector from the collision point. For the charge-conjugate decay, the unit vector becomes \mbox{$\hat{n} = -(\vec{p}_{\rm beam} \times \vec{p}_{\Lbbar})/|\vec{p}_{\rm beam} \times \vec{p}_{\Lbbar}|$}. To account for the small beam-crossing angle in LHC collisions, each particle's four-momentum is boosted into the $pp$ center-of-mass frame before calculating $\hat{n}$. 

A subsequent boost to the $\Lb$ rest frame determines $\theta$. Because $\hat{n}$ is perpendicular to the $\Lb$ flight direction, it remains unchanged under this boost. In the $\Lb$ rest frame, let $\hat n$ be the chosen polarization axis and $\hat q$ the $\Lambda$ flight direction, so that $\cos\theta=\hat n\cdot\hat q$.  
The angles $\theta_{b}$ ($\theta_{l}$) and $\phi_b$ ($\phi_l$) are the polar and azimuthal angles of the proton ($\mu^+$) in the $\Lam$ ($\Jpsi$) rest frame.
The polar angles $\theta$, $\theta_l$ and $\theta_b$ lie in $[0,\pi]$, while the azimuthal angles $\phi_l$ and $\phi_b$ lie in $[-\pi,+\pi)$.

For convenience, the coordinate system in the $\Lb$ rest frame is chosen with the $z$ axis along the $\Lambda$ flight direction and $\hat{n}$ in the $xOz$ plane:
\begin{equation}
\hat{z}_1=\hat q,\qquad
\hat{x}_1=\frac{\hat n - \hat q \cos\theta}{\sin\theta},\qquad
\hat{y}_1=\hat{z}_1 \times\hat{x}_1=\frac{\hat q\times\hat n}{\sin\theta}.
\label{eq:axes}
\end{equation}

Throughout this paper, the labels $(j,r)=(-\helicity_{\Jpsi},\helicity_\Lam)$ denote spin projections onto the $\hat{z}_1$ direction. A rotation $U_J=\exp(-i\pi J_{x_1})$ connects the $j$ and $\helicity_{\Jpsi}$ bases with $U_J|j\rangle=-|-j\rangle =-|\helicity\rangle$ for spin one; the common minus sign can be absorbed in the overall amplitude phase. 

For an incoherently produced $\Lb$ with polarization $P_b$, the production density matrix takes the form $\rho^{\Lb} = (1+P_b)/2 | +\tfrac12 \rangle\langle +\tfrac12| + (1-P_b)/2 | -\tfrac12 \rangle\langle -\tfrac12|$. A time-evolution operator $S$ in Fock space maps the initial $\Lb$ state to the $\Jpsi\Lam$ two-body state. According to the helicity formalism~\cite{jacobwick1959,1507},
\begin{equation}
|\Phi_{m_b}(\theta,\phi)\rangle=\sum_{j,r}\mathcal H_{j,r}\,
D^{\frac12}_{m_b,\ j+r}(\phi,\theta,0)^{*}|j,r\rangle  = S | m_b \rangle.
\label{eq:M}
\end{equation}
Here, we use the Wigner-$D$ functions $D^s_{mm'}(\phi,\theta,\gamma)=e^{-im\phi}d^s_{mm'}(\theta)e^{-im'\gamma}$. For a spin-half particle,
\begin{equation}
d^{\frac12}(\theta)=\begin{pmatrix}\cos\frac\theta2 & -\sin\frac\theta2\\ \sin\frac\theta2 & \cos\frac\theta2\end{pmatrix}.
\label{eq:wigner}
\end{equation}
Since the $\Lb$ particle is produced incoherently, the azimuthal angle $\phi$ is redundant and drops out.
The second Wigner index $m'=j+r$ is constrained by $|m'|\le\tfrac12$, and the $|m'|=\tfrac32$ matrix elements are therefore absent.
The couplings $\mathcal H$ are complex constants associated with the daughter spin labels in the chosen frames and do not depend on $m_b$. 
The coherence of the two-body spin states becomes observable through the subsequent decays into $\mu^+ \mu^-$ and $p\pi^-$.

The two-body ensemble density operator is $\rho_\mathrm{ens}(\theta)=S\rho^{\Lb} S^\dagger$. 
Equation (1) of Ref.~\cite{Blake:2017une} gives the full angular distribution, with the invariant mass of the lepton pair fixed to the known $\Jpsi$ mass.

Expanding Eq. (\ref{eq:M}) gives
\begin{equation}
\begin{aligned}
  \rho_\mathrm{ens}(\theta) 
  = &\frac{1+P_b}{2} | \Phi_{+\frac12}(\theta) \rangle\langle \Phi_{+\frac12} (\theta) | \\
  &+ \frac{1-P_b}{2}  | \Phi_{-\frac12}(\theta) \rangle\langle \Phi_{-\frac12} (\theta) | ,
\end{aligned}
\end{equation}
where

\begin{equation}
\begin{aligned}
|\Phi_{+\frac12}(\theta)\rangle
&=\cos\tfrac\theta2
\Bigl[\mathcal H_{0,+\frac12}\,|0,+\tfrac12\rangle
+\mathcal H_{+1,-\frac12}\,|+1,-\tfrac12\rangle\Bigr]\\
&\qquad-\sin\tfrac\theta2
\Bigl[\mathcal H_{0,-\frac12}\,|0,-\tfrac12\rangle
+\mathcal H_{-1,+\frac12}\,|-1,+\tfrac12\rangle\Bigr],
\\[6pt]
|\Phi_{-\frac12}(\theta)\rangle
&=\sin\tfrac\theta2
\Bigl[\mathcal H_{0,+\frac12}\,|0,+\tfrac12\rangle
+\mathcal H_{+1,-\frac12}\,|+1,-\tfrac12\rangle\Bigr]\\
&\qquad+\cos\tfrac\theta2
\Bigl[\mathcal H_{0,-\frac12}\,|0,-\tfrac12\rangle
+\mathcal H_{-1,+\frac12}\,|-1,+\tfrac12\rangle\Bigr].
\end{aligned}
\end{equation}

In the amplitude notation of Ref.~\cite{lhcb_jpsilam}, $\mathcal H_{0,+\frac12}\to \ap$, $\mathcal H_{0,-\frac12} \to \am$, $\mathcal H_{+1,-\frac12} \to \bp$, and $\mathcal H_{-1,+\frac12} \to \bmn$, up to a common normalization factor $\Sigma = |\ap|^2+|\am|^2+|\bp|^2+|\bmn|^2$. 
The explicit matrix elements then read
\begin{equation}
    \rho_\mathrm{ens}=\frac{1}{2\Sigma}\begin{pmatrix}
0&0&0&0&0&0\\[1pt]
0 & c_1|\bp|^2 & c_1\bp\ap^{\ast} & -x_s\,\bp\am^{\ast} & -x_s\,\bp\bmn^{\ast} & 0\\[2pt]
0 & c_1\ap\bp^{\ast} & c_1|\ap|^2 & -x_s\,\ap\am^{\ast} & -x_s\,\ap\bmn^{\ast} & 0\\[2pt]
0 & -x_s\,\am\bp^{\ast} & -x_s\,\am\ap^{\ast} & c_2|\am|^2 & c_2\am\bmn^{\ast} & 0\\[2pt]
0 & -x_s\,\bmn\bp^{\ast} & -x_s\,\bmn\ap^{\ast} & c_2\bmn\am^{\ast} & c_2|\bmn|^2 & 0\\[2pt]
0&0&0&0&0&0
\end{pmatrix},
\end{equation}
where $c_1\equiv1+P_b\cos\theta$, $c_2\equiv1-P_b\cos\theta$ and $x_s\equiv P_b\sin\theta$.
Here the ordered product basis is ($|{+},\uparrow\rangle$, $|{+},\downarrow\rangle$, $|0,\uparrow\rangle$, $|0,\downarrow\rangle$, $|{-},\uparrow\rangle$, $|{-},\downarrow\rangle$), with the $J/\psi$ factor first.
The matrix satisfies the normalization condition
\begin{equation}
  \int_{0}^{\pi} \Tr\rho_\mathrm{ens}(\theta)\,\sin\theta\,\dd\theta = 1.
\end{equation}

\section{Exact separability criterion for the \texorpdfstring{$2\otimes3$}{2x3} system}
\label{sec:ent}
We now establish exactly when the $\Jpsi\Lam$ pair is entangled. Because the system is $2\otimes3$, the Peres--Horodecki (PPT) criterion is not only necessary but also sufficient for separability~\cite{peres1996,horodecki1996}: a bipartite state is separable if and only if its partial transpose is positive semidefinite. For general $3\otimes3$ states, separability implies PPT but the reverse implication need not hold; applications to restricted $H\to VV$ families require their own support and amplitude assumptions~\cite{prd_hzz,prd_hww}.

For the general density operator $\rho_\mathrm{ens}(\theta)$ of Sec.~\ref{sec:system}, 
the partial transpose with respect to the $\Lam$ subsystem is
\begin{equation}
\bigl(\rho_\mathrm{ens}^{T_\Lam}\bigr)_{i\mu,j\nu}=\bigl(\rho_\mathrm{ens}\bigr)_{i\nu,j\mu},
\label{eq:ptdef}
\end{equation}

Here $i,j$ denote the $J/\psi$ spin labels, and $\mu,\nu$ the $\Lam$ spin labels. The state is separable if and only if the Hermitian matrix $\rho_\mathrm{ens}^{T_\Lam}$ is positive semidefinite.  

For $\Sigma>0$, $|P_b|\le1$ and $\Tr\rho_\mathrm{ens}(\theta)>0$, the state is separable if and only if
\begin{equation}
c_1\ap\bp=0,\qquad c_2\am\bmn=0,\qquad \mathrm{and}\quad x_s\bp\bmn=0.
\label{eq:ent2gen}
\end{equation}
To establish sufficiency, first consider $x_s=0$. The state is then a mixture of the two branch projectors with weights $c_1$ and $c_2$. Under the first two conditions, every nonzero-weight branch is a product ket, while a zero-weight branch makes no contribution.
If $x_s\neq0$, the three products can vanish only on the following branches:
\begin{equation}
\begin{array}{ccl}
a_+=b_-=0&:&\rho=\rho_J\otimes|\downarrow\rangle\langle\downarrow|,\\
b_+=a_-=0&:&\rho=\rho_J\otimes|\uparrow\rangle\langle\uparrow|,\\
b_+=b_-=0&:&\rho=|0\rangle\langle0|\otimes\rho_\Lambda.
\end{array}
\label{eq:product_branches}
\end{equation}
In each case, the density matrix is a tensor product, or a sum of tensor products, of two density matrices. The state is therefore separable, establishing sufficiency.

Necessity follows from the principal-minor criterion: a Hermitian matrix is positive semidefinite if and only if all its principal minors are non-negative. Among the $2\times2$ principal minors of $\rho_\mathrm{ens}^{T_\Lam}$, only three can be negative:
\begin{equation}
-\Big|\frac{c_1}{2\Sigma}\,\bp\ap^{\ast}\Big|^2,\quad
-\Big|\frac{c_2}{2\Sigma}\,\am\bmn^{\ast}\Big|^2,\quad
-\Big|\frac{x_s}{2\Sigma}\,\bp\bmn^{\ast}\Big|^2,
\label{eq:Tminors}
\end{equation}
the determinants of the principal submatrices on the subspaces spanned by
\mbox{$\{|{+},\uparrow\rangle,|0,\downarrow\rangle\}$},
\mbox{$\{|0,\uparrow\rangle,|{-},\downarrow\rangle\}$}, and
\mbox{$\{|{+},\uparrow\rangle,|{-},\downarrow\rangle\}$}, respectively.
The other $2\times2$ principal minors either vanish or are non-negative. Positive semidefiniteness of the partially transposed matrix, and hence separability, requires these three minors to be non-negative. Equation~\eqref{eq:Tminors} shows that they are also non-positive, so all three must vanish. This gives Eq.~\eqref{eq:ent2gen} and establishes necessity.

Experimental analyses often use the density matrix averaged over an angular interval. For $\theta\in[\theta_1,\theta_2]$ ($0\le\theta_1<\theta_2\le\pi$), this matrix is 
\begin{equation}
    \tilde{\rho} = \frac{\int_{\theta_1}^{\theta_2} \rho_\mathrm{ens}(\theta)\sin\theta\,\dd\theta}{\Tr\int_{\theta_1}^{\theta_2} \rho_\mathrm{ens}(\theta)\sin\theta\,\dd\theta},
\label{eq:average-rho}
\end{equation}
normalized as $\Tr\tilde{\rho}=1$. The coefficients $c_1$, $c_2$ and $x_s$ are then replaced by their averages over the range, obtained by integrating with $\sin\theta\,\dd\theta$. Because the averaged matrix has the same algebraic form with these averages in place of the pointwise weights, the criterion above applies to the binned state unchanged.

In the case $P_b=0$, the exact separability criterion reduces to
\begin{equation}
\ap\bp=0,\qquad \mathrm{and}\quad \am\bmn=0,
\end{equation}
which is independent of the angle $\theta$.

\section{Irreducible-tensor parametrization of the density operator}
\label{sec:num}
\subsection{Irreducible-tensor operators}
In terms of the irreducible-tensor operators the expansion reads
\begin{equation}
\begin{aligned}
\rho_{\mathrm{ens}}(\theta)=\frac{1}{6}\Big[&\Tr\rho_{\mathrm{ens}}(\theta)\,I_3\otimes I_2+A^{J}_{LM}\,T^{L}_{M}\otimes I_2\\
  &+A^{\Lambda}_{q}\,I_3\otimes T'_{q}\\
  &+C_{LM;q}\,T^{L}_{M}\otimes T'_{q}\Big],
\end{aligned}
\label{eq:sdm}
\end{equation}
where $T^L_M$ ($L=1,2$) are spin-one spherical tensors acting in the spin-one space, and $T'_q$ ($q=-1,0,1$) are spin-$\tfrac12$ spherical tensors acting in the two-dimensional spin-$\tfrac12$ space, using the conventions of Eq.~\eqref{eq:tensor_convention}. For the spin-$\tfrac12$ family, $T'_q=\sigma_q$ are the spherical components of the Pauli operators, with $\sigma_{\pm1}=\mp(\sigma_x\pm i\sigma_y)/\sqrt2$ and $\sigma_0=\sigma_z$. The rank-zero operator is the identity, and all other operators are traceless.
\begin{align}
T^1_{+1}&=-\sqrt{3/2}(|+\rangle\langle0|+|0\rangle\langle-|),\nonumber\\
T^2_{+1}&=-\sqrt{3/2}(|+\rangle\langle0|-|0\rangle\langle-|),\nonumber\\
T^1_0&=\sqrt{3/2}\,\mathrm{diag}(1,0,-1),\nonumber\\
T^2_0&=\mathrm{diag}(1,-2,1)/\sqrt2,\nonumber\\
T^2_{+2}&=\sqrt3|+\rangle\langle-|,\nonumber\\
T'_{+1}&=-\sqrt2\,|\!\uparrow\rangle\langle\downarrow\!|,\nonumber\\
T'_0&=|\!\uparrow\rangle\langle\uparrow\!|-|\!\downarrow\rangle\langle\downarrow\!|,\nonumber\\
T'_{-1}&=\sqrt2\,|\!\downarrow\rangle\langle\uparrow\!|.
\label{eq:tensor_convention}
\end{align}
These matrices satisfy 
$T^L_{-M}=(-1)^M(T^L_M)^\dagger$ and
$T'_{-q}=(-1)^q(T'_q)^\dagger$.
Their traces satisfy $\Tr[(T^L_M)^\dagger T^{L'}_{M'}]=3\delta_{LL'}\delta_{MM'}$ and $\Tr[(T'_q)^\dagger T'_{q'}]=2\delta_{qq'}$. The coefficients $A^{J}_{LM}$, $A^{\Lambda}_{q}$, and $C_{LM;q}$ are obtained by projecting the ensemble density matrix onto the corresponding tensor operators.
\subsection{Angular moments and the ensemble PDF}
\label{sec:ensemble_pdf}
With the angles and spin frames of Sec.~\ref{sec:system}, the normalized five-dimensional distribution is
\begin{equation}
f(\vec\Omega\mid a_\pm, b_\pm,P_b,\alpha_\Lambda)
=\frac{3}{32\pi^2}\sum_{i\in\mathcal I}M_i f_i(\vec\Omega),
\label{eq:toy_pdf}
\end{equation}
where $\alpha_\Lambda$ denotes the parity-violating asymmetry parameter of the $\Lam\to p\pi^-$ decay, $\mathcal I$ indexes the twenty nonzero angular terms, and the moments are $M_i=J_i/\Sigma$, where $J_i$ and $f_i$ are the angular coefficients and functions in Table~1 of Ref.~\cite{lhcb_jpsilam}, and $2M_1+M_2=1$. Section~\ref{sec:system} defines the full angular rate $\Sigma$. The measure, coefficient functions and moment projection weights follow Refs.~\cite{Hrivnac:1994jx,Blake:2017une,lhcb_jpsilam}, with the projection weights given by Eqs.~(4)--(5) of Ref.~\cite{lhcb_jpsilam}.

A zero-polarization sample does not completely determine the four amplitudes. After removing the common normalization and phase, or equivalently fixing one amplitude, the $P_b$-independent moments constrain the three magnitudes and two relative phases. The remaining relative phase is not uniquely determined in this limit~\cite{ATLAS:2014swk}, because the amplitudes follow from a quadratic equation with two admissible roots. Determining this phase requires a fit in which the polarization is either nonzero or allowed to vary~\cite{lhcb_jpsilam}.

\subsection{Tensor coefficients from the angular moments}
\label{sec:jdict}

The tensor expansion in Eq.~\eqref{eq:sdm} and the angular decomposition in Eq.~\eqref{eq:toy_pdf} parametrize the same rate. Each tensor coefficient is therefore a fixed function of the twenty normalized angular moments $M_i=J_i/\Sigma$ in Table~1 of Ref.~\cite{lhcb_jpsilam}, with the angular dependence entering through $\cos\theta$ and $\sin\theta$. A term-by-term matching expresses the identity component of Eq.~\eqref{eq:sdm} as
\begin{equation}
\begin{aligned}
\Tr\rho_{\mathrm{ens}}(\theta)&=\frac{2M_1+M_2+\cos\theta\,(2M_{11}+M_{12})}{2}\\
&=\frac{1+\cos\theta\,(2M_{11}+M_{12})}{2},
\label{eq:p_moments}
\end{aligned}
\end{equation}
where $2M_1+M_2=1$ and $2M_{11}+M_{12}=P_b(|\ap|^2+|\bp|^2-|\am|^2-|\bmn|^2)/\Sigma$. Thus Eq.~\eqref{eq:p_moments} reproduces the $\Tr\rho_{\mathrm{ens}}(\theta)$ of Sec.~\ref{sec:system} in terms of the measured moments.

Six of the twenty independent coefficients vanish identically,
\begin{equation}
\begin{aligned}
A^{J}_{2,2}=C_{1,0;1}&=C_{1,1;1}=C_{2,1;1}\\
&=C_{2,2;0}=C_{2,2;1}=0.
\end{aligned}
\label{eq:jzeros}
\end{equation}
These are precisely the components whose tensor operators connect the two forbidden states $|{+},\uparrow\rangle$ and $|{-},\downarrow\rangle$ to the remaining basis states. In the branch structure of Sec.~\ref{sec:ent}, the first row and last column of $\rho$ vanish identically, forcing the six projections above to zero.

Equations~\eqref{eq:jdictA}--\eqref{eq:jdictB} express the remaining independent coefficients in terms of the moments. The unlisted tensor components follow through the Hermiticity relations in Eq.~\eqref{eq:tensor_convention}. This correspondence allows the tensor coefficients, and hence the separability criteria of Sec.~\ref{sec:ent}, to be determined directly from the measured angular moments.

\begin{align}
A^{J}_{1,0}&=-\frac{\sqrt6}{2\alpha_\Lambda}\left(M_5+\cos\theta\,M_{15}\right),\label{eq:jdictA}\\[2pt]
A^{J}_{2,0}&=\sqrt2\left[(M_2-M_1)+\cos\theta\,(M_{12}-M_{11})\right],\nonumber\\[2pt]
A^{\Lambda}_{1,0}&=\frac{(2M_4+M_5)+\cos\theta\,(2M_{14}+M_{15})}{2\alpha_\Lambda},\nonumber\\[2pt]
C_{1,0;0}&=-\frac{\sqrt6}{2}\left(M_2+\cos\theta\,M_{12}\right),\nonumber\\[2pt]
C_{2,0;0}&=-\frac{\sqrt2}{\alpha_\Lambda}\left[(M_4-M_5)+\cos\theta\,(M_{14}-M_{15})\right].\nonumber
\end{align}

\begin{align}
A^{J}_{2,1}&=\frac{\sqrt3}{2}\,\sin\theta\,(M_{23}+iM_{21})=-C_{1,1;0},\label{eq:jdictC}\\[2pt]
A^{J}_{1,1}&=-\frac{\sqrt3}{2\alpha_\Lambda}\,\sin\theta\,(M_{27}+iM_{25})=-C_{2,1;0},\nonumber\\[2pt]
A^{\Lambda}_{1,1}&=\frac{1}{\sqrt2\,\alpha_\Lambda}\,\sin\theta\,(iM_{30}-M_{32})=-\frac{1}{\sqrt2}\,C_{2,0;1}.\nonumber
\end{align}

\begin{align}
C_{2,1;-1}&=\frac{\sqrt6}{2\alpha_\Lambda}\left[(M_7+iM_9)+\cos\theta\,(M_{17}+iM_{19})\right],\label{eq:jdictB}\\[2pt]
C_{2,2;-1}&=\frac{\sqrt6}{\alpha_\Lambda}\,\sin\theta\,(M_{33}+iM_{34}),\nonumber\\[2pt]
C_{1,1;-1}&=\frac{\sqrt6}{2\alpha_\Lambda}\left[\frac{M_{17}+iM_{19}}{P_b}+P_b\cos\theta\,(M_7+iM_9)\right].\nonumber
\end{align}

The expression for $C_{1,1;-1}$ holds for $P_b\neq0$. At $P_b=0$, the moments $J_{17}$ and $J_{19}$ vanish identically, so this expression no longer determines the coefficient. Its magnitude is nevertheless fixed by the $P_b$-independent moments, while its phase remains undetermined:
\begin{equation}
\begin{split}
\left|C_{1,+1;-1}\right|=\frac{\sqrt6}{2}\Bigl[&\Bigl(|\mathcal H_{0,+\frac12}|^{2}|\mathcal H_{+1,-\frac12}|^{2}\\
&+|\mathcal H_{0,-\frac12}|^{2}|\mathcal H_{-1,+\frac12}|^{2}\Bigr)\\
&-\frac{1}{\alpha_\Lambda^{2}}\left(M_7^{2}+M_9^{2}\right)\Bigr]^{1/2}.
\end{split}
\label{eq:c11_magnitude}
\end{equation}
\subsection{Entanglement carriers}
In the unpolarized family, the transverse correlations mixing $T^L_{\pm1}$ with $T'_{\pm1}$ encode the two products $\ap\bp^{\ast}$ and $\am\bmn^{\ast}$ that enter the PPT entanglement criterion. With $\Tr\rho_{\mathrm{ens}}(\theta)=\frac12$ in this family, the transverse correlation coefficients are
\begin{equation}
\begin{aligned}
C_{1,+1;-1}&=-\frac{\sqrt3}{2\Sigma}\left(\ap^{\ast}\bp+\am\bmn^{\ast}\right),\\[3pt]
C_{2,+1;-1}&=-\frac{\sqrt3}{2\Sigma}\left(\ap^{\ast}\bp-\am\bmn^{\ast}\right).
\end{aligned}
\label{eq:corr_ent}
\end{equation}
Thus $C_{1,+1;-1}=C_{2,+1;-1}=0$ if and only if $\ap\bp^{\ast}=\am\bmn^{\ast}=0$, which is equivalent to separability in this family by the PPT analysis of Sec.~\ref{sec:ent}.

For $P_b\neq0$, these two correlations keep their role but acquire the branch weights $c_1,c_2$ and involve $M_{17},M_{19}$, the $\cos\theta$ counterparts of the azimuthal pair $M_7,M_9$. A third correlation, which vanishes identically at $P_b=0$, also appears:
\begin{equation}
C_{2,2;-1}=\frac{\sqrt6}{\alpha_\Lambda}\,\sin\theta\,(M_{33}+iM_{34}).
\label{eq:corr_pol}
\end{equation}
Using the coefficients in Eq.~\eqref{eq:jdictB}, the exact condition of Sec.~\ref{sec:ent} requires all three correlations to vanish:
\begin{equation}
C_{1,+1;-1}=0,\qquad C_{2,+1;-1}=0,\qquad C_{2,2;-1}=0,
\label{eq:corr_criterion}
\end{equation}
and is equivalent to $c_1a_+b_+=c_2a_-b_-=x_sb_+b_-=0$. The weights cannot be omitted. In particular, $C_{2,2;-1}\neq0$ directly implies a nonzero transverse $\Lambda$-side product $x_sb_+b_-^{\ast}$, a correlation with no counterpart in the unpolarized family.

\section{Sensitivity comparison and experimental prospects}
\label{sec:angular}
The LHCb measurements~\cite{lhcb_jpsilam} suggest that $a_+$ and $b_-$ are suppressed, while $a_-$ and $b_+$ dominate. This hierarchy motivates the reference (null) hypothesis
\begin{equation*}
\mathcal H_0=\{a_+=b_-=0,\quad |a_-|^2+|b_+|^2>0\},
\label{eq:null_family}
\end{equation*}
with the two dominant amplitudes left free. The alternative is the complete four-amplitude family $\mathcal H_{\rm all}$.

This hypothesis provides a natural target because rejecting it has a direct entanglement interpretation. The public amplitude analysis favors $a_-\neq0$ and $b_+\neq0$, and in that domain the exact criterion of Eq.~\eqref{eq:ent2gen} becomes
\begin{equation*}
\rho\ \text{separable}\iff a_+=b_-=0 .
\end{equation*}
Equation~\eqref{eq:ent2gen} shows that any other separable configuration requires $a_-=0$ or $b_+=0$, contrary to the measured amplitude hierarchy. Thus $\mathcal H_0$ is the unique separable branch consistent with the data, and rejecting it establishes entanglement.

Several statistics for rejecting this hypothesis are compared in terms of their rejection power, defined as the probability of rejecting a false null hypothesis at a chosen significance. The two fitting methods, based on the profiled likelihood ratio and the fitted negativity, use the four-amplitude model. The third method directly uses the angular moments $M_7$ and $M_9$ without an amplitude fit. The fitting statistics are evaluated using both the full four-amplitude model and the reference two-amplitude model, and all three statistics are compared on the same pseudoexperiments. These samples are generated at a point fitted to the public LHCb moments, using the corresponding 13~TeV polarization. 

\begin{figure*}[t]
\centering
\includegraphics[width=\textwidth]{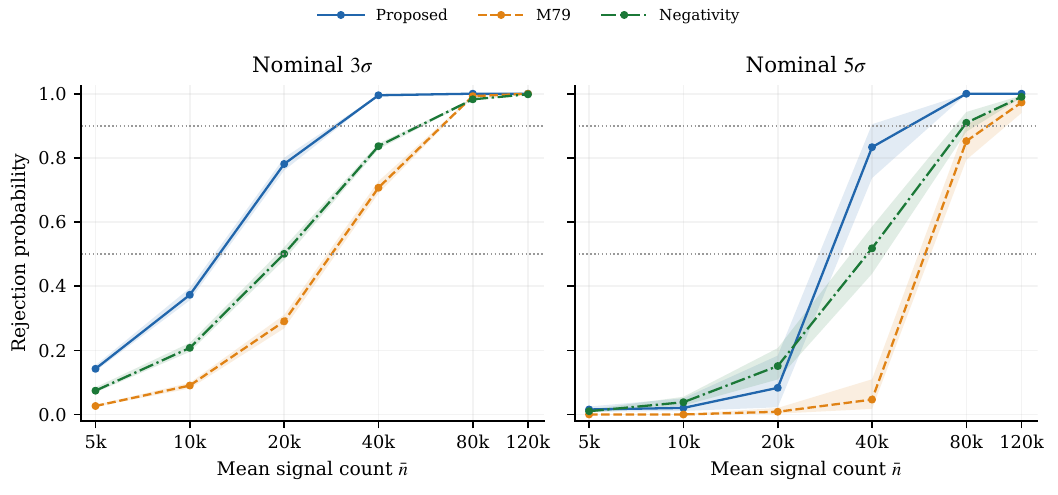}
\caption{Comparison of Negativity, M79 and Proposed: power to reject the specified $a_+=b_-=0$ null at nominal one-sided $3\sigma$ (left) and $5\sigma$ (right), for the joint 13~TeV generating point. Each count uses 10000 null and 10000 full-model toys; $P_b$ floats in both hypothesis fits. All three methods shown use generalized-Pareto null tails. Lines connect the original scan points without smoothing. Shading gives 68\% paired-bootstrap intervals. The horizontal dotted lines mark 50\% and 90\% power. The $5\sigma$ results are extrapolated projections.}
\label{fig:toy_power}
\end{figure*}
\subsection{Fitting methods}
\label{sec:test_design}
\label{sec:route2}

\emph{Profiled likelihood ratio.}
The first statistic is the profiled angular-likelihood ratio between $\mathcal{H}_0$ and $\mathcal{H}_\mathrm{all}$, with each likelihood maximized over its allowed parameters. This is the test proposed in this work, labeled \emph{Proposed} in Fig.~\ref{fig:toy_power} and Tables~\ref{tab:toy_power} and~\ref{tab:toy_targets}:
\begin{equation}
q_0=-2\ln\frac{\mathcal L(\vartheta_0,\widehat{\nu}_{\vartheta_0})}
{\mathcal L(\widehat{\vartheta},\widehat{\nu})},
\label{eq:qtilde}
\end{equation}
where $\vartheta$ denotes $(a_+, b_-)$, and $\nu$ denotes any nuisance parameters in an experimental likelihood. Both fits use the full angular distribution and optimize all allowed parameters, including $P_b$. Under $\mathcal H_0$, the free parameters include the surviving complex amplitudes $\am$ and $\bp$. The statistic therefore keeps the joint angular-shape constraints of the four-amplitude family. 

\emph{Negativity.}
The negativity~\cite{vidal2002} quantifies the entanglement of the averaged density matrix through its partial transpose:
\begin{equation}
\mathcal N=\frac12\left(\Tr\left|\tilde\rho^{T_\Lam}\right|-1\right),
\label{eq:negativity_def}
\end{equation}
where $T_\Lam$ denotes the partial transpose on the $\Lam$ factor, as defined in Sec.~\ref{sec:ent}. At $P_b=0$, the density operator is independent of $\theta$, so $\tilde\rho$ equals the conditional spin state itself and no averaging is needed. Within the four-amplitude family, its negativity has the closed form
\begin{equation}
\begin{aligned}
\mathcal N_0=\frac{1}{2\Sigma}\Big[&\sqrt{|\am|^4+4|\ap\bp|^2}-|\am|^2\\
&+\sqrt{|\ap|^4+4|\am\bmn|^2}-|\ap|^2\Big].
\end{aligned}
\label{eq:negativity0}
\end{equation}

\subsection{Non-fitting method (M7 \& M9)}
This statistic uses only the angular moments and requires no amplitude fit. For an ideal unweighted sample containing $n$ events, the moment estimators and their covariance are
\begin{equation}
\begin{aligned}
\widehat M_i&=\frac1n\sum_{c=1}^n g_i(\vec\Omega),\\
\widehat V_{ij}&=\frac{\sum_{c=1}^n(g_i(\vec\Omega)-\widehat M_i)
 (g_j(\vec\Omega)-\widehat M_j)}{n(n-1)},
\end{aligned}
\label{eq:toy_moment_covariance}
\end{equation}
Here $g_i$ are weighting functions dual to the angular basis, fixed by $\int f_j(\vec\Omega)\,g_i(\vec\Omega)\,\dd\vec\Omega=(32\pi^2/3)\delta_{ij}$. Requiring them to lie in the span of the $f_j$ makes this choice unique. The two moments used below have $g_7=\tfrac{15}{2}f_7$ and $g_9=\tfrac{15}{2}f_9$; Refs.~\cite{Beaujean:2015xea,Blake:2017une} give the complete set. In experimental data, the same averages are formed with event weights that correct for background and for nonuniform detector acceptance and selection efficiency~\cite{lhcb_jpsilam}. 

The angular moments 
\begin{equation}
M_7 = \frac{\alpha_\Lambda}{\sqrt{2}\Sigma} \mathrm{Re}(-b_+^* a_+ + b_- a_-^*),\qquad M_9 = \frac{\alpha_\Lambda}{\sqrt{2}\Sigma} \mathrm{Im}(b_+^* a_+ - b_- a_-^*)
\end{equation}
are sensitive to the separation criteria.
A nonzero value of either moment is sufficient to indicate entanglement.

The statistic is the projected interference test of Eq.~\eqref{eq:toy_T79},
\begin{equation}
T_{79}=\widehat{\mathbf m}^{\mathsf T}V^{-1}\widehat{\mathbf m},
\qquad \widehat{\mathbf m}=(\widehat M_7,\widehat M_9)^{\mathsf T},
\label{eq:toy_T79}
\end{equation}
with $V$ the covariance of the moment pair estimated in each toy. The statistic preserves only the two-dimensional information in these projected interference moments, without requiring a preliminary four-amplitude fit. Its zero-moment set is wider than $\mathcal H_0$ because it also includes configurations in which the two interference contributions cancel.

Within the unpolarized four-amplitude family and under the stated analyzer assumptions, nonzero interference is a sufficient but not necessary signature of entanglement. Vanishing $M_7,M_9$, or equivalently $C_{2,1;-1}=0$, does not imply separability because $C_{1,1;-1}$ may remain nonzero. For polarized conditional states, the direction-dependent criterion remains Eqs.~\eqref{eq:ent2gen} and~\eqref{eq:corr_criterion}.

\subsection{Pseudoexperiment configurations}
\label{sec:common_toys}
The statistics are applied to a common set of idealized samples with the same generated events, full-model and null generating points, event-count settings and fixed decay-analyzer input. A joint fit to the public moments supplies the full generating point and its 13~TeV polarization, while a constrained joint fit supplies the corresponding null point. Each toy has a Poisson-fluctuated signal count. The comparison uses the mean signal counts
\begin{equation*}
\bar n\in\{5000,\,10000,\,20000,\,40000,\,80000,\,120000\},
\label{eq:toy_counts}
\end{equation*}
which correspond to the nodes in Fig.~\ref{fig:toy_power}. The toys assume unit efficiency, no backgrounds and no resolution smearing. Appendix~\ref{sec:toy_sensitivity} documents the full parameters, sample counts and fit checks.

Each statistic is calibrated against its own null distribution at the same one-sided significance. This toy calibration is distinct from the Gaussian-moment calibration of the public M79 result.

The generated samples do not populate the $5\sigma$ tail sufficiently to estimate the rejection power by direct counting, so the null tails must be extrapolated. Above a threshold $u$, set to the 90th percentile of the null sample, each null distribution is modeled with a generalized Pareto distribution whose survival function is
\begin{equation}
S(T)=r\left[1+\xi\frac{T-u}{\beta}\right]^{-1/\xi},
\qquad T\geq u,
\label{eq:toy_tail}
\end{equation}
on the support where the bracket is positive. Here $\beta>0$ is the scale and $r$ the empirical exceedance fraction; the limit $\xi=0$ gives the exponential survival function~\cite{coles2001}. The parameters are fitted by maximum likelihood to the exceedances using the density $p(T)=-\dd S/\dd T$. The magnitude of the negativity, the mean value of a statistic and its fitted uncertainty provide related diagnostics, but the comparison below uses the rejection power itself.

\begin{table*}[t]
\centering
\caption{Rejection probabilities (\%) for the joint 13~TeV generating point. Each entry is $3\sigma\,/\,5\sigma$ and uses the same 10000 full-generated toys at that mean count. All three methods use generalized-Pareto null tails. The $5\sigma$ results are extrapolations. Values of 100\% denote the observed toy fraction.}
\label{tab:toy_power}
\setlength{\tabcolsep}{18pt}
\begin{tabular}{@{}rccc@{}}
\toprule
$\bar n$ & Proposed & M79 & Negativity \\
\midrule
5000 & 14.26 / 1.58 & 2.66 / 0.03 & 7.44 / 1.04 \\
10000 & 37.31 / 2.06 & 9.04 / 0.02 & 20.78 / 3.84 \\
20000 & 78.09 / 8.32 & 29.05 / 0.87 & 50.09 / 15.11 \\
40000 & 99.51 / 83.33 & 70.69 / 4.66 & 83.65 / 51.80 \\
80000 & 100.00 / 100.00 & 99.24 / 85.22 & 98.25 / 90.93 \\
120000 & 100.00 / 100.00 & 99.98 / 97.28 & 99.80 / 99.02 \\
\bottomrule
\end{tabular}
\end{table*}

\begin{table*}[t]
\centering
\caption{Approximate mean ideal signal counts for 50\% and 90\% power, obtained by interpolating the probit of the power linearly in the square root of the mean signal count between adjacent scan points. Before the probit transform, powers are clipped to $[0.5/10001,\,1-0.5/10001]$. Counts are rounded to the nearest hundred. Threshold calibration follows the prescriptions of Sec.~\ref{sec:common_toys}; the $5\sigma$ targets depend on the extrapolated null tails.}
\label{tab:toy_targets}
\begin{tabular*}{\textwidth}{@{\extracolsep{\fill}}lrrrr}
\toprule
 & \multicolumn{2}{c}{$3\sigma$} & \multicolumn{2}{c}{$5\sigma$} \\
Method & 50\% power & 90\% power & 50\% power & 90\% power \\
\midrule
Proposed & 12600 & 24900 & 30900 & 43600 \\
M79 & 29200 & 54000 & 63000 & 89900 \\
Negativity & 20000 & 49300 & 39000 & 78000 \\
\bottomrule
\end{tabular*}
\end{table*}
\subsection{Comparative sensitivity and prospects}
\label{sec:comparison}
Figure~\ref{fig:toy_power} shows the central comparison, and Table~\ref{tab:toy_targets} lists the event targets. The main results are summarized here, with numerical details deferred to Appendix~\ref{sec:toy_sensitivity}.

At fixed significance and power, the proposed likelihood method reaches the evidence-level ($3\sigma$) and nominal observation-level ($5\sigma$) targets with fewer ideal signal events than either M79 or fitted negativity. The gain reflects the amount of angular information retained by each statistic. Fitted negativity compresses the full fit into one scalar, and an entanglement measure optimized for other purposes need not maximize the power against this particular hypothesis. M79 uses two projected moments, discarding the joint constraints on amplitudes, populations and polarization encoded in the full angular distribution; interference cancellation can further reduce its response. 

In these idealized pseudoexperiments, the profiled likelihood test reaches the target powers with fewer signal events than the other two statistics. It therefore offers a promising route to evidence-level sensitivity with samples on the existing scale and to nominal observation-level sensitivity with larger samples. The required ideal event scale is $\mathcal O(10^4)$, comparable to the signal samples in the published angular analysis~\cite{lhcb_jpsilam}. The larger Run~3 data set provides further motivation~\cite{lhcb_run3_2025}; Appendix~\ref{app:sample_scale} compares the available sample sizes.

These projections nevertheless rely on idealized pseudoexperiments with a fixed generating point, unit efficiency, no backgrounds and no resolution smearing. A detector-level implementation must account for acceptance, backgrounds, resolution and nuisance parameters, including the uncertainty in the generating description. 

\section{Information from the existing angular analysis}
\label{sec:data}
\begin{figure}[tb]
\centering
\includegraphics[width=\singlefigwidth]{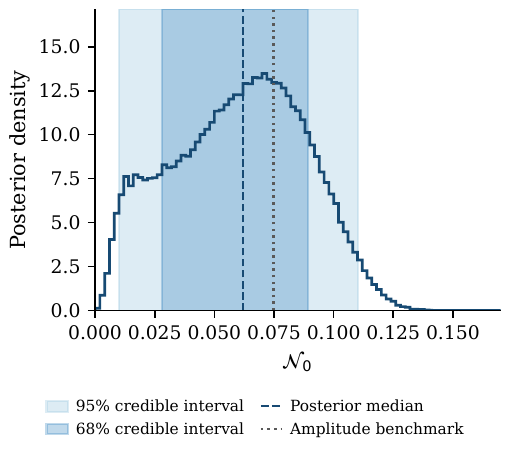}
\caption{Posterior negativity of the unit-trace unpolarized-reference conditional state reconstructed from the public LHCb moments, including statistical and systematic uncertainties and the published prior prescription. Shading indicates the equal-tailed 68\% and 95\% intervals. The dotted line is the negativity evaluated at the marginal posterior modes quoted by LHCb~\cite{lhcb_jpsilam}, and the dashed line is the posterior median of this nonlinear observable.}
\label{fig:negativity_posterior}
\end{figure}
The public angular analysis of $\Lb\to\Jpsi\Lam$ provides the common experimental input for the reconstructions and tests above~\cite{lhcb_jpsilam}. Its published results directly constrain several entanglement-related observables.

\subsection{Negativity}
\label{sec:posterior}
\label{sec:modeldep}
Equation~\eqref{eq:negativity_def}, evaluated at the most probable values of the four amplitudes reported in Ref.~\cite{lhcb_jpsilam}, gives the benchmark negativity
\begin{equation*}
    \mathcal{N}_0 = 0.075.
\end{equation*}
A posterior distribution of the negativity is also estimated. 
The public-moment likelihood, the experimental covariance, and the prior prescription are from Ref.~\cite{lhcb_jpsilam}; sampling the resulting joint amplitude posterior and evaluating Eq.~\eqref{eq:negativity0} for each sample gives the negativity posterior. The common amplitudes and decay-analyzer parameter are inferred jointly across $pp$ center-of-mass energies. Because the measured samples do not have $P_b$ fixed to zero, the separate production polarizations are marginalized in this likelihood.

Figure~\ref{fig:negativity_posterior} shows the inferred magnitude and uncertainty, with a posterior median of approximately $0.062$. This posterior propagates the parameter uncertainty from the existing data. By contrast, the fitted-negativity statistic introduced in Sec.~\ref{sec:test_design} is evaluated once per toy experiment and calibrated as a hypothesis test. Under continuous amplitude priors, the exactly separable subfamilies have zero prior measure. A positive lower credible bound therefore cannot, by itself, be interpreted as a frequentist exclusion significance.

\subsection{The joint \texorpdfstring{$M_7$--$M_9$}{M7--M9} test}
\label{sec:modelindep}

Under a common-moment assumption, we combine the moments measured at 7, 8 and 13~TeV in Ref.~\cite{lhcb_jpsilam} together with their covariance. This gives an M79 rejection of $M_7=M_9=0$ at approximately $2.75\sigma$, including statistical and systematic uncertainties. This result uses the Gaussian-moment approximation. At each energy, the $2\times2$ block of $(\widehat M_7,\widehat M_9)$ in that covariance is augmented by adding the squared systematic errors to its diagonal. The moment pairs are then combined across energies with inverse-variance weights under the common-moment assumption.
Under the zero-moment null, the observable $T_{79}$ of Eq.~\eqref{eq:toy_T79} follows a $\chi^2_2$ distribution, giving $p=\exp(-T_{79}/2)$. The statistical-only result is $2.80\sigma$, while the 13~TeV sample alone gives $1.95\sigma$. Appendix~\ref{app:numbers} records the combined moments and the consistency check across energies.

The existing moment-based result therefore lies close to the usual evidence threshold. This motivates examining whether the full angular likelihood can improve the rejection power for a related, physically motivated reference hypothesis.

\section{Summary}
\label{sec:concl}

We have analyzed entanglement in the $2\otimes3$ spin system formed by the $J/\psi$ and $\Lambda$ in \mbox{$\Lb\to\Jpsi(\to\mu^+\mu^-)\,\Lam(\to p\pi^-)$} decays, using a common spin basis and decay-analyzer frames. Within the four-amplitude family $\{a_\pm,b_\pm\}$, the helicity amplitudes determine the spin-density matrix of the pair, for which the Peres--Horodecki test yields an exact separability criterion. The entanglement can then be expressed directly in terms of measured angular moments. These results have been applied to the published LHCb angular analysis, and three procedures for a prospective interference measurement have been compared.

Within the specified spin basis and four-amplitude family, without any $CP$ or amplitude-symmetry assumption, the exact criterion is $\ap\bp=\am\bmn=0$ at $P_b=0$, and $c_1\ap\bp=c_2\am\bmn=x_s\bp\bmn=0$ for arbitrary physical polarization and nonzero conditional rate ($c_{1,2}=1\pm P_b\cos\theta$, $x_s=P_b\sin\theta$). The additional $x_s\bp\bmn$ condition removes the purely transverse branch from the separable set and follows from the forbidden-state principal minors.  Written in the irreducible-tensor form of Sec.~\ref{sec:num}, the same criterion is the vanishing of the three transverse correlations $C_{1,+1;-1}=C_{2,+1;-1}=C_{2,2;-1}=0$; at $P_b=0$ only the first two survive, and all three are fixed by the measured angular moments.

The public angular moments constrain both the amount of entanglement within this family and departures from zero interference. The posterior reconstruction gives a median negativity $\mathcal N\simeq0.06$ (with a $68\%$ credible interval of about $[0.03,0.09]$) for the unpolarized-reference conditional state, while the combined-data M79 test gives approximately $2.75\sigma$ rejection of zero interference in the Gaussian-moment approximation. These are distinct inferences from the existing data, and the latter already approaches the usual evidence threshold.

For the selected $a_+=b_-=0$ reference hypothesis, common pseudoexperiments compare Negativity, M79 and the profiled angular-likelihood test using the same significance and power criteria. At $50\%$ power, the profiled test requires about $1.3\times10^4$ events for $3\sigma$ and $3.1\times10^4$ for $5\sigma$, approximately $51$--$57\%$ fewer than M79 across the studied targets. When the dominant $a_-$ and $b_+$ amplitudes are nonzero, this hypothesis is the only separable branch. Its rejection therefore directly probes entanglement within the experimentally favored hierarchy, and the LHCb Run~2 signal samples already exceed the $3\sigma$ target. These prospects call for dedicated measurements at the LHC experiments, where the practical complexity of angular acceptance, backgrounds, resolution and nuisance parameters must be carefully addressed.

\begin{acknowledgments}
This work was supported by the National Key R\&D Program of China under contract No. 2024YFA1610500.
\end{acknowledgments}

\appendix
\section{Details of the common pseudoexperiments}
\label{sec:toy_sensitivity}
The test statistics and common comparison criteria are defined in Sec.~\ref{sec:test_design}. This appendix records the generating points, the numerical results and the comparison with available sample scales.

The angular measure used throughout is
\begin{equation}
\begin{gathered}
\dd\vec\Omega=\dd\cos\theta\,\dd\cos\theta_l\,\dd\phi_l\,
\dd\cos\theta_b\,\dd\phi_b,\\
\int\dd\vec\Omega\,\frac{3}{32\pi^2}\sum_iJ_if_i=2J_1+J_2=\Sigma,
\end{gathered}
\label{eq:angular_measure}
\end{equation}
with all three cosines in $[-1,1]$ and both azimuths in $[-\pi,\pi)$. Since $2J_1+J_2=\Sigma$ and $2M_1+M_2=1$, the normalized PDF of Eq.~\eqref{eq:toy_pdf} integrates to unity without additional factor.

\subsection{Generating points and ensembles}
\label{sec:joint_point}
For the prospective sensitivity study, the Gaussian-moment likelihood of Eq.~(6) of Ref.~\cite{lhcb_jpsilam} is minimized over the common complex amplitudes and $\alpha_\Lambda$, together with separate $P_b$ values at 7, 8 and 13~TeV. This fit retains the full statistical covariance and uses the same systematic-error treatment as the posterior reconstruction. The generating amplitudes are the coordinates of one joint optimum. By contrast, the marginal posterior modes quoted by LHCb are separate one-dimensional modes and need not form a joint mode. With the convention that $b_+=1$ is real, the result is
\begin{equation}
\begin{aligned}
a_+&=0.111736e^{2.17922i},&a_-&=1.025294e^{1.07537i},\\
b_+&=1,&b_-&=0.169515e^{2.44411i}.
\end{aligned}
\label{eq:joint_benchmark}
\end{equation}
The fitted polarizations are $(-0.00188,-0.01846,0.05435)$ at the three energies, and $\alpha_\Lambda=0.7540757$. The minimum quadratic value is $38.715$ for 57 moments plus the external constraint and ten fitted parameters. These amplitudes give $\Sigma=2.092449$ and an unpolarized-reference conditional negativity of $\mathcal N_0=0.0857452$. This point value differs from the posterior median in Sec.~\ref{sec:posterior} and does not represent the direction-dependent negativity at nonzero $P_b$.

The full-model toys use the 13~TeV value $P_b=0.0543456$. A separate fit to the public moments, constrained to $a_+=b_-=0$, supplies the null generating amplitudes and polarization specified below. Both five-dimensional PDFs use the same fixed $\alpha_\Lambda=0.7540757$. The posterior uncertainty reconstruction in Sec.~\ref{sec:posterior} and the choice of a single generating point serve different purposes: the latter defines a conditional sensitivity projection without propagating the uncertainty of the generating parameters.

The null generating point is obtained from this constrained fit:
\begin{equation}
\begin{aligned}
a_+&=b_-=0,\qquad b_+=1,\\
a_-&=1.020307e^{0.99003i},\qquad P_b=0.0484226,\\
t_0&=\frac{|a_-|^2}{1+|a_-|^2}=0.5100503.
\end{aligned}
\label{eq:toy_null_fraction}
\end{equation}
Although the constrained fit gives $\alpha_\Lambda=0.7528918$, the toy comparison uses the common full-fit value $0.7540757$ in both generators and all fits, while keeping the constrained null amplitudes and $P_b$ above. Discrimination therefore does not arise from different decay-analyzer inputs. Both hypothesis fits allow $P_b\in[-1,1]$ to vary. The toys assume unit efficiency, no backgrounds and no resolution smearing. The full polarized angular PDF is given by Eq.~\eqref{eq:toy_pdf}, with the angular functions in Table~1 of Ref.~\cite{lhcb_jpsilam}. All tests in Sec.~\ref{sec:test_design} use the same events. The common Poisson yield provides no discrimination between the hypotheses, which are distinguished only through their angular shapes.

Event generation uses uniform proposals over the five-dimensional angular domain, accepted according to the full polarized PDF with a bound obtained from the sum of the absolute angular-term bounds. The generator evaluates the original LHCb coefficient and angular-function implementation. Although the generated events depend on the joint-fit parameters, they contain no measured LHCb events.

For each toy, the $(7,9)$ block of Eq.~\eqref{eq:toy_moment_covariance} supplies the M79 covariance. Repeated starts, convergence checks and comparisons with the original angular-PDF implementation are used to assess both amplitude fits. The fixed overall phase and total scale leave the allowed relative phases free, including those relevant at nonzero polarization.

\subsection{Numerical results}
\label{app:toy_calibration}
The proposed test has greater sensitivity than M79 at the studied power targets, as shown in Fig.~\ref{fig:toy_power} and Table~\ref{tab:toy_power}. At $\bar n=40000$, its $3\sigma$ and $5\sigma$ powers are 99.51\% and 83.33\%, compared with 70.69\% and 4.66\% for M79. The paired power gains are 28.82 and 78.67 percentage points, with 68\% bootstrap intervals $[26.76,30.64]$ and $[65.95,85.34]$ percentage points, respectively. As a check independent of the tail functional form, direct empirical $3\sigma$ calibration gives 99.50\% versus 68.78\% at the same event count.

Interpolating the power curves gives the approximate event targets in Table~\ref{tab:toy_targets}. At 50\% power, the proposed test requires about 12600 events for $3\sigma$ and 30900 for $5\sigma$, compared with 29200 and 63000 for M79. At 90\% power, the corresponding counts are about 24900 and 43600 versus 54000 and 89900. Across these targets, the likelihood method uses approximately 51--57\% fewer ideal signal events because it retains the joint angular-shape constraints of the amplitude family beyond the two projected azimuthal moments.

The fitted-negativity test also requires larger event counts at these power targets. At $\bar n=40000$, its $3\sigma$/$5\sigma$ powers are 83.65\%/51.80\%. This comparison concerns the particular scalar statistic specified above and does not imply a universal ordering of negativity-based tests, or even of all extrapolated powers at individual scan points. For example, a likelihood ratio with an equivalent constraint can retain the full likelihood information.

As a tail-model cross-check, fitting a gamma distribution to the profiled null statistic gives 50\%/90\% power targets of approximately 30600/47000 events at $5\sigma$, while a $\chi^2_2$ tail for $T_{79}$ gives 67600/104800. These alternative approximations preserve the event-count advantage but demonstrate that the extreme-tail projections depend on the calibration choice.

\subsection{Comparison with available sample scales}
\label{app:sample_scale}
The available LHCb Run~2 signal samples are already on this scale. Summing the long and downstream categories, Table~2 of Ref.~\cite{lhcb_jpsilam} reports 2647 and 19100 signal decays in 2015 and 2016, respectively. Their total of 21747 exceeds the ideal 50\%-power target for $3\sigma$ but falls below that for $5\sigma$. The remainder of Run~2 and the larger Run~3 data set offer further opportunities; LHCb recorded $11.8\,\mathrm{fb}^{-1}$ of high-quality $pp$ data in 2025 alone~\cite{lhcb_run3_2025}. These samples motivate an experimental study, although the achievable significance will depend on angular acceptance, backgrounds and control of nuisance parameters.

\section{Numerical record}
\label{app:numbers}
This appendix collects the numbers quoted in the text, using the methods of Ref.~\cite{lhcb_jpsilam} and Secs.~\ref{sec:test_design} and~\ref{sec:posterior}.

\subsection{Posterior of the negativity}
For the public LHCb moments, the posterior of the unpolarized-reference negativity has
\begin{equation}
\begin{aligned}
\operatorname{median}(\mathcal N_0)&=0.06185,\\
\mathcal N_0&\in[0.02813,\,0.08911]\quad(68\%),\\
\mathcal N_0&\in[0.00993,\,0.11016]\quad(95\%),
\end{aligned}
\label{eq:negativity_posterior}
\end{equation}
with equal-tailed credible intervals. The posterior mean and standard deviation are $0.06015$ and $0.02770$. Sixteen chains provide $1.28\times10^6$ retained, correlated samples after warm-up; the effective sample size of the negativity is approximately $1.79\times10^4$ and $\widehat R=1.0016$. Reweighting to a uniform prior on the normalized population simplex gives a median of $0.07445$ and a 68\% interval $[0.04637,0.09839]$, which illustrates the prior dependence near the amplitude boundary.

\subsection{Marginal-mode benchmark}
The marginal posterior modes quoted by LHCb~\cite{lhcb_jpsilam}, written with $b_+=1$ real, are
\begin{equation}
\begin{aligned}
a_+&=0.129e^{-2.523i},& a_-&=1.021e^{1.122i},\\
b_+&=1,& b_-&=0.145e^{1.788i}.
\end{aligned}
\label{eq:benchmark}
\end{equation}
These amplitudes give $\Sigma=2.080107$ and $\mathcal N_0=0.0748444$. At $P_b=0.5$, they give $\mathcal N(0)=0.06154$, $\mathcal N(\pi/2)=0.08310$ and $\mathcal N(\pi)=0.10487$. The rate-weighted mean using the $\theta$-dependent rate $N(\theta)$ is $0.0830529$, while uniform averages over $\theta$ and $\cos\theta$ give $0.0829959$ and $0.0829696$, respectively. Although all three round to $0.083$, they represent different measures. Independent variation of the moduli within the quoted ranges, with $|b_+|=1$, spans $\mathcal N_0\in[0.03013,0.09758]$; this box is not a joint confidence or credible region.

\subsection{Public $M_7$--$M_9$ combination}
The combined moments are $\widehat M_7=0.02610\pm0.00986$ and $\widehat M_9=-0.02102\pm0.00988$, with correlation coefficient $0.00827$. This gives $T_{79}=11.6229$, $p=2.9931\times10^{-3}$ and $Z=2.75$; the statistical-only significance is $2.80\sigma$. The 13~TeV sample alone gives $T_{79}=7.3454$, $p=0.02541$ and $Z=1.95$, while the 7 and 8~TeV samples give $T_{79}=3.3543$ and $2.2851$ ($Z=0.89$ and $0.47$). The common-moment fit is consistent across energies, with $\chi^2=1.3619$ for four degrees of freedom ($p=0.851$).

\subsection{Moments at the generating point}
For the generating point of Eq.~\eqref{eq:joint_benchmark} and $\alpha_\Lambda=0.7540757$,
\begin{equation}
\begin{aligned}
J_7&=0.0526511,& J_9&=-0.0419010,\\
M_7&=0.0251624,& M_9&=-0.0200248.
\end{aligned}
\label{eq:J79num}
\end{equation}
The interference ratio is $|\am^{\ast}\bmn-\ap\bp^{\ast}|/(|\am\bmn^{\ast}|+|\ap\bp^{\ast}|)=0.44196$. Inserting instead the marginal modes of Eq.~\eqref{eq:benchmark}, with $\alpha_\Lambda=0.750$, gives $(M_7,M_9)=(0.0564730,-0.0423923)$ and a ratio of $0.99972$.

\input{Lb_JpsiLambda_v3.bbl}

\end{document}

%% file: styles/lhcb-like.tex
\usepackage[top=1in, bottom=1.25in, left=1in, right=1in]{geometry}
\newcommand{\singlefigwidth}{0.5\textwidth}

\usepackage{microtype}
\usepackage[font=small,labelfont=small]{caption}
\usepackage[bottom,flushmargin,hang,multiple]{footmisc}
\makeatletter
\g@addto@macro\bfseries{\boldmath} %
\makeatother

\usepackage{cite}
\usepackage{mciteplus}
\usepackage{ifthen}
\newboolean{articletitles}\setboolean{articletitles}{true}   %
\newboolean{pdflatex}\setboolean{pdflatex}{true}
\newboolean{uprightparticles}\setboolean{uprightparticles}{false}

\makeatletter
\def\lhcb@authors{}
\def\author#1{%
  \ifx\lhcb@authors\@empty
    \g@addto@macro\lhcb@authors{#1}%
  \else
    \g@addto@macro\lhcb@authors{,\quad#1}%
  \fi}
\def\lhcb@affil{}
\def\affiliation#1{%
  \ifx\lhcb@affil\@empty
    \g@addto@macro\lhcb@affil{#1}%
  \else
    \g@addto@macro\lhcb@affil{\\#1}%
  \fi}
\def\lhcb@mails{}
\def\email{\@ifnextchar[{\lhcb@email}{\lhcb@email[]}}
\def\lhcb@email[#1]#2{%
  \ifx\lhcb@mails\@empty
    \g@addto@macro\lhcb@mails{\texttt{#2}}%
  \else
    \g@addto@macro\lhcb@mails{,\ \texttt{#2}}%
  \fi}

\newsavebox{\lhcb@abstractbox}
\renewenvironment{abstract}{%
  \global\setbox\lhcb@abstractbox\vbox\bgroup
    \hsize=\textwidth \small
    \leftskip=\leftmargini \rightskip=\leftmargini
    \noindent\ignorespaces
}{%
  \par\egroup
}

\renewcommand\maketitle{%
  \hypersetup{pdftitle={\@title}}%
  \pagenumbering{roman}%
  \renewcommand{\thefootnote}{\fnsymbol{footnote}}%
  \begin{titlepage}%
    \vspace*{2cm}%
    {\centering
      {\LARGE\bfseries \@title\par}%
      \vspace{2.0cm}%
      {\large \lhcb@authors$^{1,*,\dagger}$\par}%
      \bigskip
      {\normalfont\itshape\footnotesize $^{1}$\lhcb@affil\par}%
      \vspace{1.2em}%
      {\normalsize \@date\par}%
    \par}%
    \vspace{\fill}%
    {\centering {\bfseries Abstract}\par}%
    \vspace{0.6em}%
    \noindent\usebox{\lhcb@abstractbox}\par
    \vspace{\fill}%
    \footnotetext[1]{These authors contributed equally to this work.}%
    \footnotetext[2]{Contact authors: \lhcb@mails.}%
  \end{titlepage}%
  \renewcommand{\thefootnote}{\arabic{footnote}}%
  \setcounter{footnote}{0}%
  \setcounter{page}{2}%
  \newpage\mbox{~}%
  \clearpage
  \pagenumbering{arabic}%
  \setcounter{page}{1}%
  \pagestyle{plain}%
}

\newenvironment{acknowledgments}{\section*{Acknowledgments}}{}

\makeatother

%% file: Lb_JpsiLambda_v3.bbl
\ifx\mcitethebibliography\mciteundefinedmacro
\PackageError{LHCb.bst}{mciteplus.sty has not been loaded}
{This bibstyle requires the use of the mciteplus package.}\fi
\providecommand{\href}[2]{#2}